\documentclass[aip,reprint,twocolumn,amsmath,amssymb]{revtex4-1}

\usepackage{graphicx}
\usepackage{dcolumn}
\usepackage{bm}

\usepackage[utf8]{inputenc}
\usepackage[T1]{fontenc}
\usepackage[english]{babel}
\usepackage{mathptmx}
\usepackage{xcolor}
\usepackage{stmaryrd}
\usepackage{soul}
\usepackage{amsmath}

\begin{document}

\title{On the Gibbs-Thomson equation for the crystallization of confined fluids} 

\author{Laura Scalfi}
\affiliation{Physicochimie des \'Electrolytes et Nanosyst\`emes Interfaciaux, Sorbonne Universit\'e, CNRS, 4 Place Jussieu F-75005 Paris, France}

\author{Beno\^it Coasne}
\affiliation{Laboratoire Interdisciplinaire de Physique, Universit\'e Grenoble Alpes, CNRS, France}

\author{Benjamin Rotenberg}
\email[]{benjamin.rotenberg@sorbonne-universite.fr}

\affiliation{Physicochimie des \'Electrolytes et Nanosyst\`emes Interfaciaux, Sorbonne Universit\'e, CNRS, 4 Place Jussieu F-75005 Paris, France}

\date{\today}

\begin{abstract}
The Gibbs-Thomson (GT) equation describes the shift of the crystallization temperature for a confined fluid with respect to the bulk as a function of pore size. While this century old relation is successfully used to analyze experiments, its derivations found in the literature often rely on nucleation theory arguments (\emph{i.e.} kinetics instead of thermodynamics) or fail to state their assumptions, therefore leading to similar but different expressions. Here, we revisit the derivation of the GT equation to clarify the system definition, corresponding thermodynamic ensemble, and assumptions made along the way. We also discuss the role of the thermodynamic conditions in the external reservoir on the final result. We then turn to numerical simulations of a model system to compute independently the various terms entering in the GT equation, and compare the predictions of the latter with the melting temperatures determined under confinement by means of hyper-parallel tempering grand canonical Monte Carlo simulations. We highlight some difficulties related to the sampling of crystallization under confinement in simulations. Overall, despite its limitations, the GT equation may provide an interesting alternative route to predict the melting temperature in large pores, using molecular simulations to evaluate the relevant quantities entering in this equation. This approach could for example be used to investigate the nanoscale capillary freezing of ionic liquids recently observed experimentally between the tip of an Atomic Force Microscope and a substrate. 
\end{abstract}

\maketitle 


\section{Introduction}

Most fluid properties are modified under confinement due to the interactions with the confining surfaces. Of particular importance is the shift of phase transitions, which is more pronounced for small pore sizes (large surface to volume ratio). Such a shift depends on the excess free energies associated with the interface between the pore walls and both coexisting phases\cite{gelb_phase_1999}. For instance, the capillary condensation of vapor inside a pore occurs at a pressure lower than the saturation pressure corresponding to the bulk liquid-vapor equilibrium, with a shift described by the Kelvin equation\cite{evans_fluids_1990}. The crystallization of confined fluids, such as in freeze-thaw cycles or salt crystallization in porous rocks and stones, is also of great practical importance to understand weathering in the context of the durability of civil engineering constructions or  the preservation of cultural heritage. 
The fact that the crystallization of a confined fluid occurs at a different temperature than in the bulk can be exploited to investigate the properties of ``supercooled'' water (even though the confinement also has an influence on these properties), or to estimate pore size distributions in complex porous materials, \emph{e.g.} via NMR-cryoporometry. Several reviews are available on the effect of confinement on freezing/melting as probed using experiments and molecular simulations are available in the literature\cite{alba-simionesco_effects_2006, alcoutlabi_effects_2005}.
 
The shift of the melting temperature $T_m$ induced by the confinement of the liquid in a slit pore of width $H$ is traditionally described by the Gibbs-Thomson (GT) equation:
\begin{equation}\label{eq::gt}
  \frac{T_{m} - T_{m}^b}{T_{m}^b} = \frac{2 \left( \gamma_{LW} - \gamma_{SW} \right)}{H \rho \Delta _{m}h} \quad ,
\end{equation}
where $T_{m}^b$ is the bulk melting temperature, $\gamma_{LW}$ and $\gamma_{SW}$ are respectively the liquid-wall and solid-wall surface tensions, $\rho = N/V$ the density and $\Delta_{m}h = h_L - h_S$ the latent heat of melting per particle. Since the latter is usually positive, for a given fluid the sign of the shift is determined by that of the surface tension difference, \emph{i.e.} the difference in the free energy cost to create an interface between each of the phases and the confining walls. This balance is often complex to predict as it is significantly system-dependent: for example, recent experiments on the capillary freezing of ionic liquids between the tip of an Atomic Force Microscope (AFM) and a solid substrate indicate that the switch to a mechanical response typical of a solid occurs at a distance which depends on the metallicity of the substrate~\cite{comtet_nanoscale_2017}.

Even though the GT equation has been used for more than a century, one finds in the literature a variety of expressions, which differ not only because they may correspond to different geometries but also in the use of the liquid or solid density in the denominator\cite{warnock_geometrical_1986,awschalom_supercooled_1987,kaneko_elevation/depression_2017,bresme_computer_2006,koga_phase_2005,christenson_confinement_2001,evans_fluids_1990,jackson_melting_1990,petrov_curvature-dependent_2006,scherer_crystallization_1999,ritter_collective_1988,nath_chakraborty_monte_2012}. In addition, its derivations do not always state explicitly the assumptions that are made at the different steps. We believe that some of the ambiguities that can be found in the literature  are due to the similarity between the thermodynamic problem of phase equilibrium under confinement, where two phases are stable, and the kinetic problem of nucleation, where one phase is more stable than the other but the growth of a nucleus is hindered by the free energy cost associated with the creation of an interface. Even if these two aspects have in common the presence of interfaces and associated surface free energies and lead to similar expressions, they correspond to different thermodynamic conditions and processes (so that the similar expressions correspond to different physical quantities).

From the nucleation point of view, one considers the kinetic barrier for the solid to grow from the liquid phase, when the former is thermodynamically more stable than the latter\cite{buffat_size_1976}. Classical nucleation theory involves the free energy associated with the interface between the two phases and the chemical potential difference between them at the considered thermodynamic conditions -- typically, fixed temperature $T$ and pressure $P$. The competition between the bulk driving force and the cost of creating the interface leads (a) to a critical nucleus size, which also reflects the curvature of the interface and satisfies a relation similar to the GT equation Eq.~\ref{eq::gt}, and (b) to the corresponding free energy barrier. This barrier controls the kinetics of the phase transition and explains why the liquid may be cooled down below the bulk melting temperature without observing crystallization. One can note in passing that the standard assumption of a spherical nucleus, which is reasonable for the liquid-vapor transitions, is questionable for the nucleation of solids, which are faceted objects (leading in addition to facet-dependent interfacial free energies)\cite{valeriani_rate_2005}.

For crystallization under confinement, arguments borrowing from this nucleation picture have been proposed to derive the shift in melting temperature induced by confinement~\cite{warnock_geometrical_1986,awschalom_supercooled_1987}.
However, the GT equation deals with the thermodynamic equilibrium between the two phases in the presence of confining walls (typically, slit or cylindrical pores). In that case, the relevant interfaces and associated free energies are not between the solid and liquid phases but between each of them and the walls. The interfacial free energies will in general differ for the confined liquid and the confined solid, so that one of them is more stable than the other at the bulk melting temperature $T_m^b$. Conversely, the melting temperature $T_m$ under confinement is shifted with respect to $T_m^b$. These considerations are not related to the formation of an interface between the two confined phases. As a result, several important simulations studies have performed free energy calculations using umbrella sampling to probe crystallization under confinement by estimating the free energy of the confined liquid and crystal phases without explicitly considering their interface\cite{hung_molecular_2005,radhakrishnan_global_2002}.

\begin{figure*}[htb!]
    \centering
    \includegraphics[width=1.\linewidth]{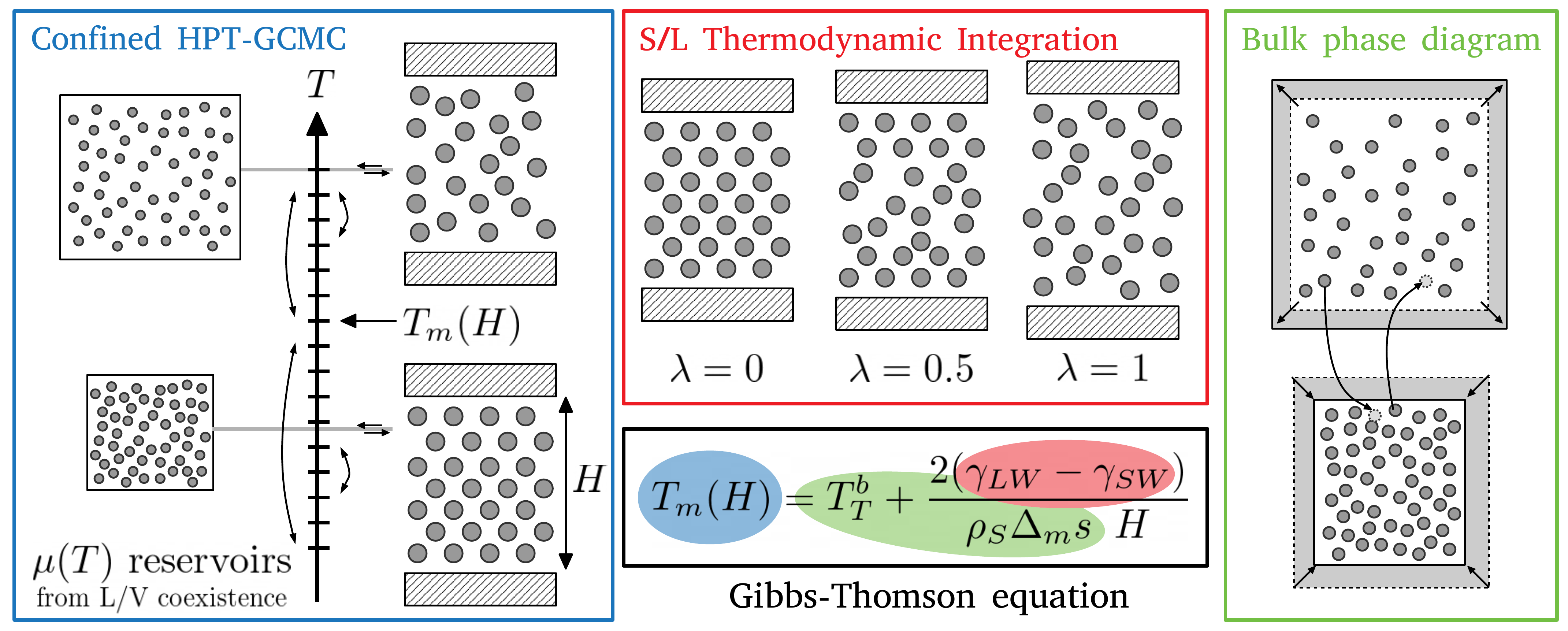}
    \caption{
    The Gibbs-Thomson equation (see Eq.~\ref{eq::gtT} below) describes the shift of the melting temperature $T_m$ due to confinement, as a function of the confining distance $H$. It involves bulk properties (melting temperature, density and melting entropy) as well as interfacial ones (difference between the liquid-wall and solid-wall surface tensions). In the present work, we estimate these terms independently and compare the prediction of the GT equation to the melting temperature obtained in hyper-parallel tempering grand-canonical Monte-Carlo simulations under confinement.
    }
    \label{fig:schema}
\end{figure*}

In the present work, we propose a derivation of the GT equation for the crystallization of a liquid confined in a slit pore, based only on the phase equilibrium of the confined phases. We discuss in particular the importance of the definition of the system and of the thermodynamic ensemble corresponding to an experimental situation. We then estimate independently the various terms entering in the GT equation for a model system and compare the prediction of this equation to the melting temperature under confinement determined in simulations.
In Section~\ref{sec:GT}, we provide a derivation of the GT equation and discuss the assumptions leading to the final result. The rest of the article is then devoted to the numerial study of the phase behaviour of a model system, introduced in Section~\ref{sec:system}, in order to test the relevance of these assumptions and of the GT equation to predict the shift of its melting temperature. This requires the computation of several quantities, using complementary strategies as schematized in Fig.~\ref{fig:schema}. The bulk phase diagram and relevant properties of the bulk phases are investigated in Section~\ref{sec:bulk}. Section~\ref{sec:TI} presents the computation of differences in interfacial free energies under confinement using a thermodynamic integration approach. Finally, Section~\ref{sec:hptgcmc} discusses crystallization under confinement by comparing results from Hyper-Parallel Tempering Grand-Canonical Monte Carlo (HPT-GCMC) simulations with the prediction of the GT equation.

\section{Deriving the Gibbs-Thomson equation under confinement}\label{sec:GT}

As mentioned in the previous section, some derivations in the literature refer to metastable states using arguments related \emph{e.g.} to supersaturation or undercooling under given thermodynamic conditions. In contrast, in the following, we consider only the equilibrium phases at coexistence and determine the coexistence line in the space of relevant thermodynamic variables. The derivation, which largely borrows from that of Evans \emph{et al.} for capillary condensation using a slightly different ensemble~\cite{evans_phase_1987, evans_fluids_1986, evans_capillary_1986, evans_fluids_1990, dominguez_monte_1999}, allows to focus on the effect of the confining walls (W) on the phase equilibrium. 
Even though we consider here the solid-liquid coexistence and a slit-like pore, it can be easily adapted to different confining geometries or conditions.
The derivation proceeds in two steps. Firstly, we identify the relevant thermodynamic ensemble and associated thermodynamic potential to derive a ``confined Clapeyron'' formula satisfied by the thermodynamic variables along the coexistence line. Secondly, integration along this line to connect the bulk conditions to the confined ones leads to the GT equation.


\subsection{A "confined Clapeyron" approach in the $\mu A_WHT$ ensemble}

\begin{figure}[hbt!]
    \centering
    \includegraphics[width=1.\linewidth]{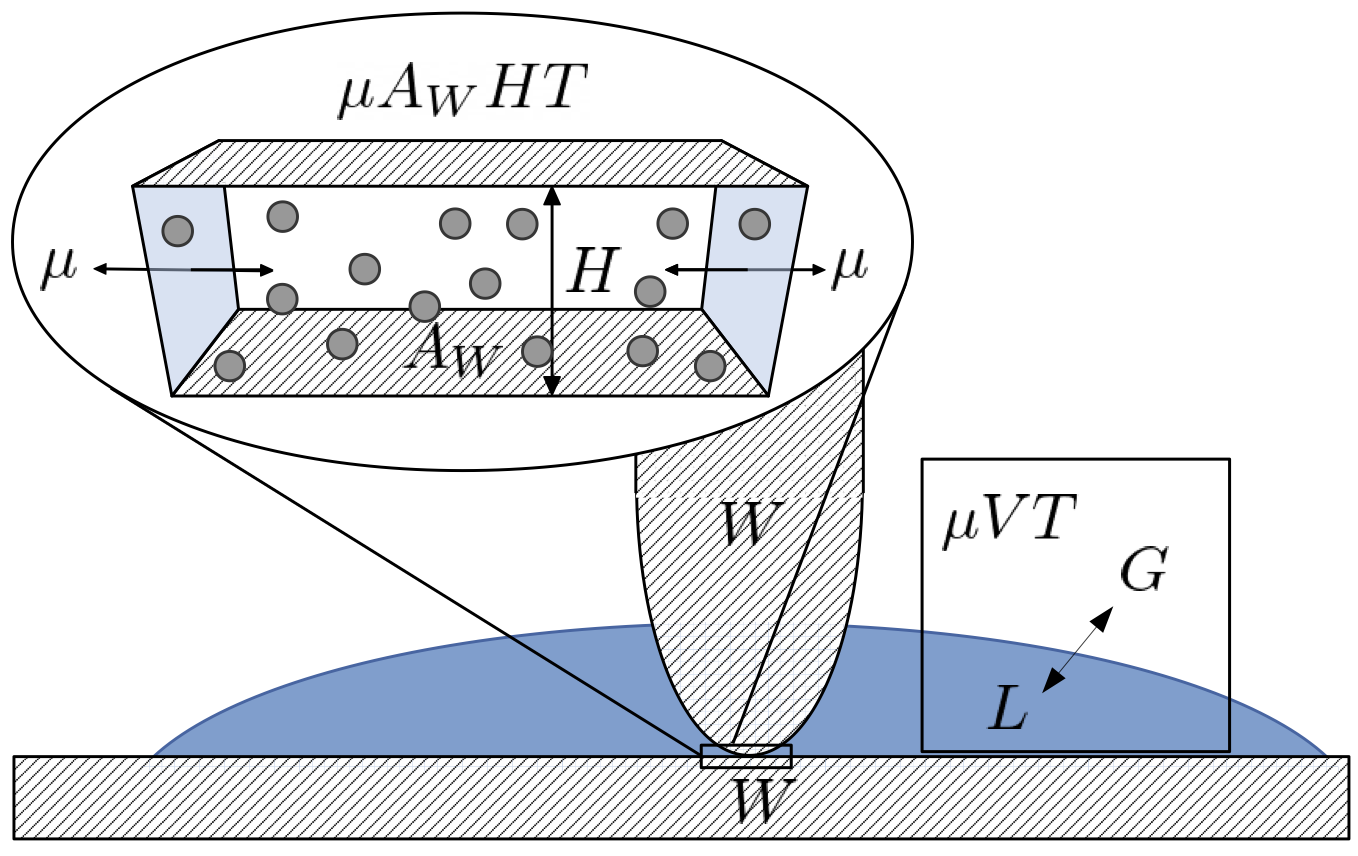}
    \caption{
    Illustration of a system confined between the tip of an Atomic Force Microscope and a substrate, as in the experiments of Ref.~\citenum{comtet_nanoscale_2017}. From the thermodynamic point of view, it forms an open system confined in a slit pore with lateral area $A_W$ and distance between walls $H$, at a fixed temperature $T$ and chemical potential $\mu$ set by the liquid-gas equilibrium in the reservoir.}
    \label{fig:gibbs-thomson}
\end{figure}

Our system of interest corresponds to the experimental setup of Ref.~\citenum{comtet_nanoscale_2017}, schematized in Fig.~\ref{fig:gibbs-thomson}, where the tip of an AFM confines a room temperature ionic liquid that undergoes capillary freezing at a finite distance $H$ between the tip and the substrate. The height at which this transition occurs depends on the nature of the substrate. At the macroscopic level, the interactions of the subtrate with the confined fluid/solid are reflected in the surface tensions, as discussed below. Because of the large radius of curvature of the tip, the region in which the phase transition occurs can be considered as a slit pore between two parallel walls. The slit pore has a surface area $A_W$, a width  $H$ and, hence,  a pore volume $V=A_WH$ (see the inset of Fig.~\ref{fig:gibbs-thomson}). In contrast to the experiments in Ref.~\citenum{comtet_nanoscale_2017}, we will assume that both confining walls are made of the same material, so that their interactions with the confined fluid or solid are identical. The rest of the liquid in which the AFM tip is placed can be considered as a macroscopic reservoir, so that the chemical potential $\mu$ is fixed and the number of particles $N$ in the confined, open system fluctuates. In addition, the whole system is maintained at a fixed temperature $T$, so that the thermodynamic ensemble corresponding to this experiment is the $\mu A_W H T$ ensemble. In the following, we consider the relevant thermodynamic variables both in the liquid and solid phases, indicated by subscripts $L$ and $S$, respectively. We emphasize that these phases are considered separately, \emph{i.e.} that there is no interface between them (unlike in nucleation-inspired approaches).

In the $\mu A_W H T$ ensemble, the thermodynamic potential is the grand potential
\begin{align}
    \Omega &= U - TS - \mu N = - P A_W H + 2 \gamma A_W
    \; ,
    \label{eq:omega}
\end{align}
with the internal energy
\begin{align}
    U &= TS - PA_W H + 2 \gamma A_W + \mu N
    \; ,
\end{align}
$S$ the entropy, $P$ the pressure and $\gamma$ the surface tension.
At coexistence between the liquid and solid phases, the thermodynamic potentials of the two phases are equal, \emph{i.e.} $\Omega_L=\Omega_S$. This is not the case of thermodynamic derivatives: introducing this last equality in Eq.~\ref{eq:omega}, it follows that the pressure in each phase differ by
\begin{equation}
P_L - P_S = \frac{2 (\gamma_{LW} - \gamma_{SW})}{H} 
\label{eq::pcpl}
\end{equation}
which depends on the difference in surface tension between the liquid and the walls, and between the solid and the walls, respectively, as well as on the pore size $H$.
We note again that this difference is not related to the presence of an interface between the two confined phases.

We now consider the changes in the grand potential associated with a change in the thermodynamic variables defining the ensemble. From the first principle of thermodynamics and the expressions of the work associated with changes in the height $H$ and surface area $A_W$, one obtains
\begin{align}
    d\Omega &= -SdT - PA_WdH + (2\gamma - PH) dA_W - Nd\mu 
    \; .
\end{align}
We then follow the reasoning of the Clausius-Clapeyron equation, which gives the slope $dP/dT$ of the coexistence line (in the $P$, $T$ plane) for a bulk system, and consider an infinitesimal change in the control variables while staying at coexistence, so that $\Omega_S + d\Omega_S = \Omega_L + d\Omega_L$ along this path, giving $d\Omega_S = d\Omega_L$. After simplification of the term in $dA_W$ using Eq.\ref{eq::pcpl}, we obtain
\begin{gather}
(S_L-S_S) dT + \frac{2 A_W (\gamma_{LW} - \gamma_{SW})}{H} dH + (N_L - N_S) d\mu = 0 
\; .
\label{eq::domega}
\end{gather}
Equation \ref{eq::domega} relates the variations of $T$, $H$ and $\mu$ along the solid-liquid coexistence. We note that, as expected, Eq.~\ref{eq::domega} shows that the confinement effect does not depend on the variations of the surface area.

\subsection{Integration along a bulk-to-confined thermodynamic path}

In order to obtain the GT equation, we will integrate Eq.~\ref{eq::domega} from an unconfined, bulk system ($H\to\infty$) where the transition occurs at the bulk melting temperature $T_m^b$, to another point along the coexistence line with a melting temperature $T_m$ for a finite distance $H$ between the confining surfaces. 
This requires introducing some additional setup-specific information on the thermodynamic conditions in the reservoir, allowing us to express the dependence of the chemical potential with the temperature $d\mu/dT$. We thus rearrange Eq.~\ref{eq::domega} as
\begin{gather}\frac{dH}{H^2} = - \frac{dT}{2 (\gamma_{LW} - \gamma_{SW})} \left[ (\rho_L s_L-\rho_S s_S) + (\rho_L -\rho_S) \frac{d\mu}{dT} \right] 
\; ,
\label{eq::dhgen}
\end{gather}
where we used the densities $\rho = N/V$ and entropies per particle $s = S/N$. Eq.~\ref{eq::dhgen} defines the L-S coexistence line under confinement by the joint variations of $H$ and $T$.
In addition, unlike in the steps leading to Eq.~\ref{eq::domega}, we will make some assumptions (discussed along the derivation and numerically in the next sections) on some physical quantities.

In the experiments of Ref.~\citenum{comtet_nanoscale_2017}, the liquid is in equilibrium with its vapour, as shown in the right part of Fig.~\ref{fig:gibbs-thomson}. The chemical potential of the reservoir in equilibrium with the confined system is fixed by the bulk liquid-gas coexistence, \emph{i.e.} $\mu(T) = \mu_L(T) = \mu_G(T)$, where the $L$ and $G$ subscripts refer to the liquid and gas phase, respectively. 
Its derivative with respect to temperature is given by (see Appendix~\ref{app:dmudt})
\begin{equation}
\label{eq:approx}
\frac{d\mu}{dT} = - \frac{\rho_L^b s_L^b-\rho_G^b s_G^b}{\rho_L^b - \rho_G^b} \approx - s_L^b \, ,
\end{equation}
where the superscript $b$ refers to the bulk (unconfined) liquid and gas phases.
In the case of an isobaric-isothermic liquid phase, the result would be exactly $-s_L^b$. Introducing Eq.~\ref{eq:approx} into Eq.~\ref{eq::dhgen}, we obtain
\begin{gather}
\frac{dH}{H^2} = - \frac{(\rho_L -\rho_S) (s_L - s_L^b)+\rho_S (s_L - s_S)}{2 (\gamma_{LW} - \gamma_{SW})} dT \, .
\label{eq::dhgen2}
\end{gather}
For sufficiently large confining distances (and corresponding small shift in the melting temperature $T_m-T_m^b$), one can approximate the densities and entropies per particle of the confined phases by their bulk counterparts. The first term in the numerator can safely be neglected (since, in addition to this assumption, $|\rho_L -\rho_S|\ll\rho_S$), and we obtain
\begin{equation}
    \frac{dH}{H^2} = - \frac{\rho_S^b \Delta_ms^b}{2 (\gamma_{LW} - \gamma_{SW})} dT 
    \, ,
\end{equation}
with $\Delta_ms^b=s_L^b - s_S^b$ the bulk entropy of melting per particle. 

The final step to recover the GT equation is to integrate this equation along a thermodynamic path connecting the confined system for a finite $H$ and corresponding $T_m$ and an unconfined one ($H\to\infty$ and bulk melting temperature $T_m^b$). To this end, we 
assume that the ratio on the right-hand side is independent of temperature and confining distance over the considered range. This approximation should be accurate at least for sufficiently large $H$ and corresponding small $T_m-T_m^b$; it will be tested numerically and discussed in section~\ref{sec:bulk}. Under these conditions, we can write
\begin{gather} \label{eq:integralGT}
    \int \limits _H ^{+\infty} \frac{dH}{H^2} = -  \frac{\rho_S^b \Delta_m s^b}{2 (\gamma_{LW} - \gamma_{SW})} \int \limits _{T_{m}} ^{T_{m}^b} dT
\end{gather}
Noting that in the present case of a liquid-gas equilibrium in the reservoir the melting temperature is in fact the (bulk) triple point $T_T^b$, the final result can be written as
\begin{equation}\label{eq::gtT}
    T_{m}(H) = T_T^b + \frac{2 (\gamma_{LW} - \gamma_{SW})}{H\rho_S^b \Delta_{m} s^b}
    \; .
\end{equation}
This derivation can be easily adapted to other geometries or external reservoir conditions. In the case where the reservoir is an isobaric liquid, the first term in the right-hand side is simply the bulk melting temperature at the corresponding pressure.


\section{Model system}\label{sec:system}

To assert the validity of the assumptions in the above derivation (in particular, neglecting the temperature dependence of some quantities), we use molecular simulation to compute the various terms entering in the GT equation for a simple system of Lennard-Jones (LJ) particles confined between unstructured walls. More precisely, in order to avoid the difficulties associated with the long-range corrections (LRC) in the computation of physical properties under confinement, we consider the truncated shifted Lennard-Jones (TSLJ) potential
for a pair of atoms $i$ and $j$ at a distance $r_{ij}$,
\begin{equation}
  u^{TSLJ}_{ij}(r_{ij}) =
    \begin{cases}
      u_{ij}(r_{ij}) - u_{ij}(r_\mathrm{cut}) & \text{if } r_{ij} < r_\mathrm{cut}\\
      0 & \text{otherwise,}
    \end{cases}       
\end{equation}
where $r_\mathrm{cut}$ is the cutoff radius, and 
\begin{equation}
u_{ij}(r_{ij}) = 4 \epsilon \left[ \left( \frac{\sigma}{r_{ij}}\right)^{12} - \left( \frac{\sigma}{r_{ij}}\right)^{6} \right]
\end{equation}
with $\epsilon$ and $\sigma$ the LJ energy and diameter.
The total energy of the system is then given by
\begin{equation}
U_{tot} = \frac{1}{2}\sum \limits_i \sum \limits_j u_{ij}^{TSLJ}(r_{ij})
\end{equation}
where the sums run over all atoms in the system. Simulations are performed with typical values for argon\cite{hansen_phase_1969}: $\epsilon = 119.8$~K and $\sigma = 3.405$~\AA. We employ a cutoff radius $r_\mathrm{cut} = 2.5 \sigma$, for which some data on the phase diagram is available in the literature.
In the following, most quantities will be given in reduced LJ units, indicated by a $^*$ superscript: $r^* = r/\sigma$ for distances, $E^* = E/\epsilon$ for energies, $T^* = k_BT/\epsilon$ for temperatures (with $k_B$ the Boltzmann constant), $P^* = P\sigma^3/\epsilon$ for pressures and $\gamma^* = \gamma \sigma^2/\epsilon$ for surface tensions.

As for the confining medium, we use the unstructured Steele wall\cite{steele_physical_1973,steele_interaction_1978}, with parameters corresponding to a mica surface\cite{cui_molecular_2001}, which is for an atom at a distance $z$ from the surface
\begin{multline}
u^\mathrm{Steele} (z) = 2 \pi \rho_W \epsilon_{WF} \sigma_{WF}^2 \Delta \left[ \frac{2}{5} \left( \frac{\sigma_{WF}}{z}\right)^{10} \right.
\\ \left. - \left( \frac{\sigma_{WF}}{z}\right)^{4} - \frac{\sigma_{WF}^4}{3 \Delta (z+0.61\Delta)^3}\right]
\end{multline}
where $\rho_{W}^* = 1.0$ is the atomic density of the material, $\epsilon_{WF}^* = \sqrt{\epsilon_{W}^*}$ and $\sigma_{WF}^* = (1 + \sigma_{W}^*)/2$ are obtained by combining LJ parameters with surface parameters $\epsilon_W^* = 7.85$ and $\sigma_W^* = 1.28$, and $\Delta^* = 0.84$, corresponds to the distance between atomic crystal planes. In practice, we use two walls separated by a distance $H$ which controls the pore size (see Section~\ref{sec:hptgcmc}).

The crystalline structure of the Lennard-Jones fluid is a face-centered cubic phase and the most stable face that crystallises on the Steele wall is the (111) face, with which subsequent simulation boxes were initialized. Because crystallization is a phenomenon particularly sensitive to the box size, we worked on crystal configurations consistent between the different techniques. Care was taken to ensure the reversibility of calculations and to avoid polycrystalline recrystallization: boxes were initialized from a perfect crystal and liquid configurations were obtained by melting. Simulated systems in section~\ref{sec:bulk} consisted of 4000 atoms per phase in a cubic box of variable volume; in section~\ref{sec:TI} of 4116 atoms with lateral dimensions $L_x^* = 16.4$, $L_y^* = 14.2$ and a pore size $H^*$ between Steele walls fluctuating between $19.5$ and $21.5$. In Section~\ref{sec:hptgcmc}, we use boxes with $L_x^* = 21.1$, $L_y^* = 20.3$ and different $H^*$ values (8.7, 11.6, 14.5, 17.3, 20.2, 23.1, 26.0 and 28.8), covering a wider range than studies focussing on disjoining pressure effects, which typically consider pores up to $\approx10$ molecular diameters.


\section{Bulk properties}\label{sec:bulk}


The phase diagram of LJ particles has been extensively studied\cite{hansen_phase_1969,ladd_triple-point_1977,agrawal_solid-fluid_1995,agrawal_thermodynamic_1995,agrawal_thermodynamic_1995-1,mastny_melting_2007,wang_lennard-jones_2020}, mostly using LRC to correct for the use of a cutoff to compute the interactions. For the TSLJ potential however, the choice of $r_\mathrm{cut}$ greatly influences the phase diagram~\cite{ahmed_effect_2010,ghoufi_computer_2016}. For example, the use of TSLJ with a cutoff of $2.5\sigma$ results in a critical temperature difference of about 35~K for an argon fluid with respect to the prediction with LRC~\cite{hansen_phase_1969}. For this cutoff value, Vrabec \emph{et al.}~\cite{vrabec_comprehensive_2006} computed the liquid-vapor coexistence line and determined the critical point, whereas Ahmed and Sadus~\cite{ahmed_effect_2010} investigated the solid-liquid coexistence line at high pressure.

In order to accurately locate the triple point, we recompute both the liquid-vapor and the solid-liquid portions of the phase diagram using Gibbs-Duhem Integration (GDI)\cite{kofke_direct_1993}. Starting from a known point on the coexistence line in the ($T,P$) plane, a new point is found by integrating the Clausius-Clapeyron equation 
\begin{equation}
\label{eq:clapeyron}
\frac{\mathrm{d}\ln P}{\mathrm{d} \beta} = - \frac{\Delta_r h}{P\beta \Delta_r v}
\; ,
\end{equation}
where $\beta = 1/k_BT$ and $\Delta_r h$ and $\Delta_r v$ are the transition enthalpy and volume per particle, respectively. The right-hand side is computed on-the-fly in simulations in the $NPT$ ensemble of two systems corresponding to the two coexisting phases, (\emph{i.e.} liquid and vapor, or liquid and solid). More details are given in Appendix~\ref{app:GDI}.

The starting point for the GDI method is crucial: one needs to accurately identify one point of the coexistence line, because the integration of Eq.~\ref{eq:clapeyron} allows to stay on the latter but not to find it.
For the liquid-vapor transition, we start from a coexistence point at high temperature ($T_m^* = 1.00, P_m^*=0.0612 \pm 0.0005$), determined using Gibbs-Ensemble Monte Carlo (GEMC) simulations\cite{panagiotopoulos_direct_1987,panagiotopoulos_phase_1988}. GEMC determines the coexistence between two phases by exchanging volume and particles between two systems, until the chemical potential is equal in the two phases. While efficient for the liquid-vapor equilibrium, GEMC is insufficient for the liquid-solid one, due to the low probability of particle exchange. Fortunately, at high pressure and temperature the difference in the phase behavior of the LJ with LRC and of the TSLJ potentials becomes negligible. We therefore use as a starting point the results of Kofke \emph{et al.} on the LRC-LJ system\cite{agrawal_thermodynamic_1995}: $T_m^* = 2.74, P_m^*=36.9$.

\begin{figure}[hbt!]
    \centering
    \includegraphics[width=1.\linewidth]{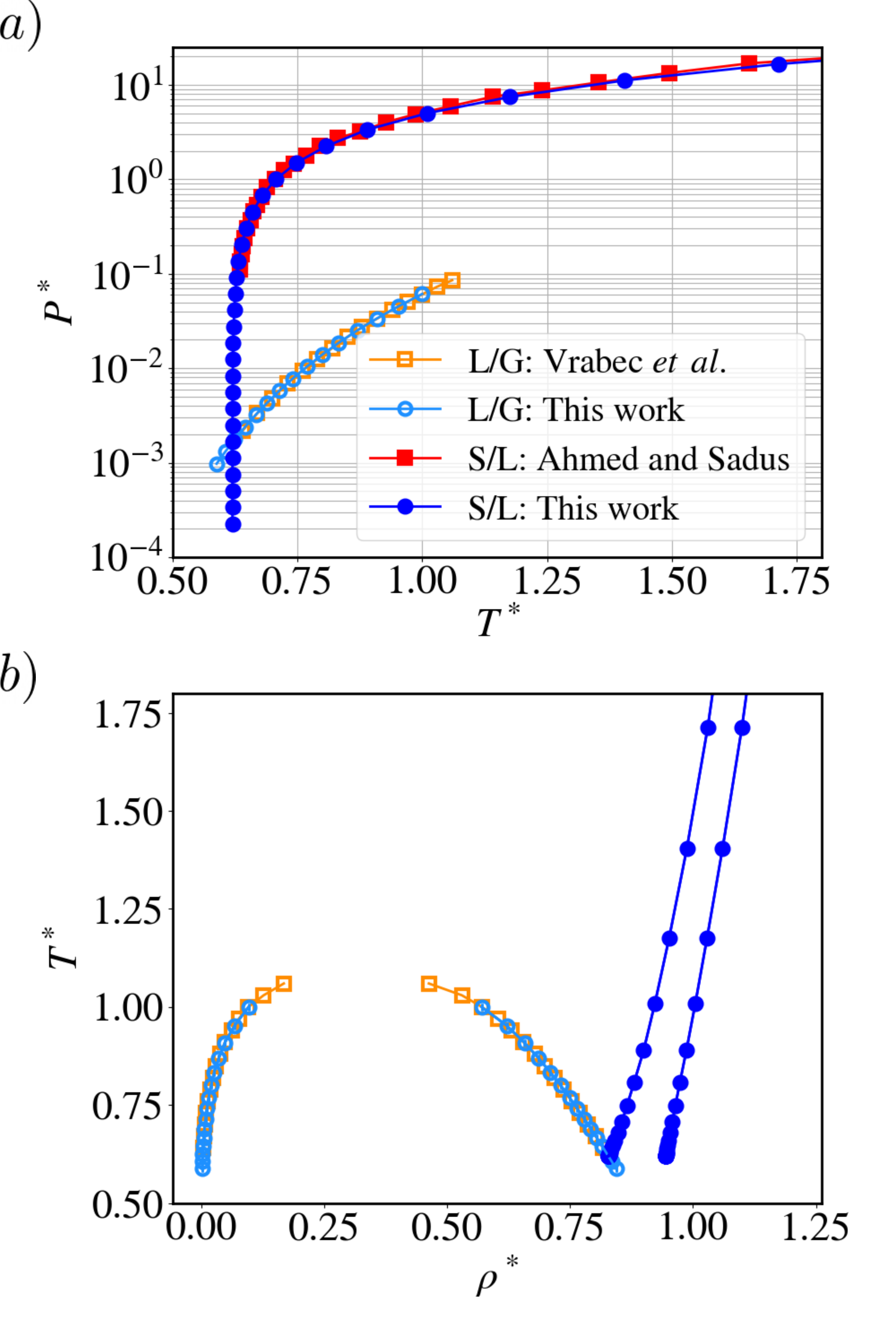}
    \caption{
    Bulk phase diagram of the truncated shifted Lennard-Jones system with a cutoff of $2.5\sigma$: (a) in the $(T^*, P^*)$ plane; (b) in the $(\rho^*, T^*)$ plane. Our results obtained by Gibbs-Duhem integration for the liquid-vapor (light blue open circles) and solid-liquid (dark blue full circles) coexistence lines are compared to the results of Vrabec \emph{et al.}~\cite{vrabec_comprehensive_2006} (orange open squares) and Ahmed and Sadus~\cite{ahmed_effect_2010} (red full squares), respectively. All thermodynamic quantities are in LJ units.
    }
    \label{fig:bulkphd}
\end{figure}

Fig.~\ref{fig:bulkphd} shows the resulting phase diagram, together with the coexistence lines from Vrabec \emph{et al.}~\cite{vrabec_comprehensive_2006} and Ahmed and Sadus~\cite{ahmed_effect_2010} for comparison. Panel~\ref{fig:bulkphd}a, in the $(T^*, P^*)$ plane, demonstrates the very good agreement with available literature data. Moreover, from our extended range of considered thermodynamic conditions we can locate the triple point for the TSLJ with a cutoff of  $2.5\sigma$, which corresponds to $T_T^* = 0.62, P_T^*=1.65~10^{-3}$. Panel~\ref{fig:bulkphd}b, in the $(\rho^*, T^*)$ plane, further shows that the density of the liquid and solid phases differ by 7-15\%, especially at lower temperatures, so that using $\rho_L^b$ instead of $\rho_S^b$ in the GT Eq.~\ref{eq::gtT} leads to a different estimate of the melting temperature under confinement.

\begin{figure}[hbt!]
    \centering
    \includegraphics[width=1.\linewidth]{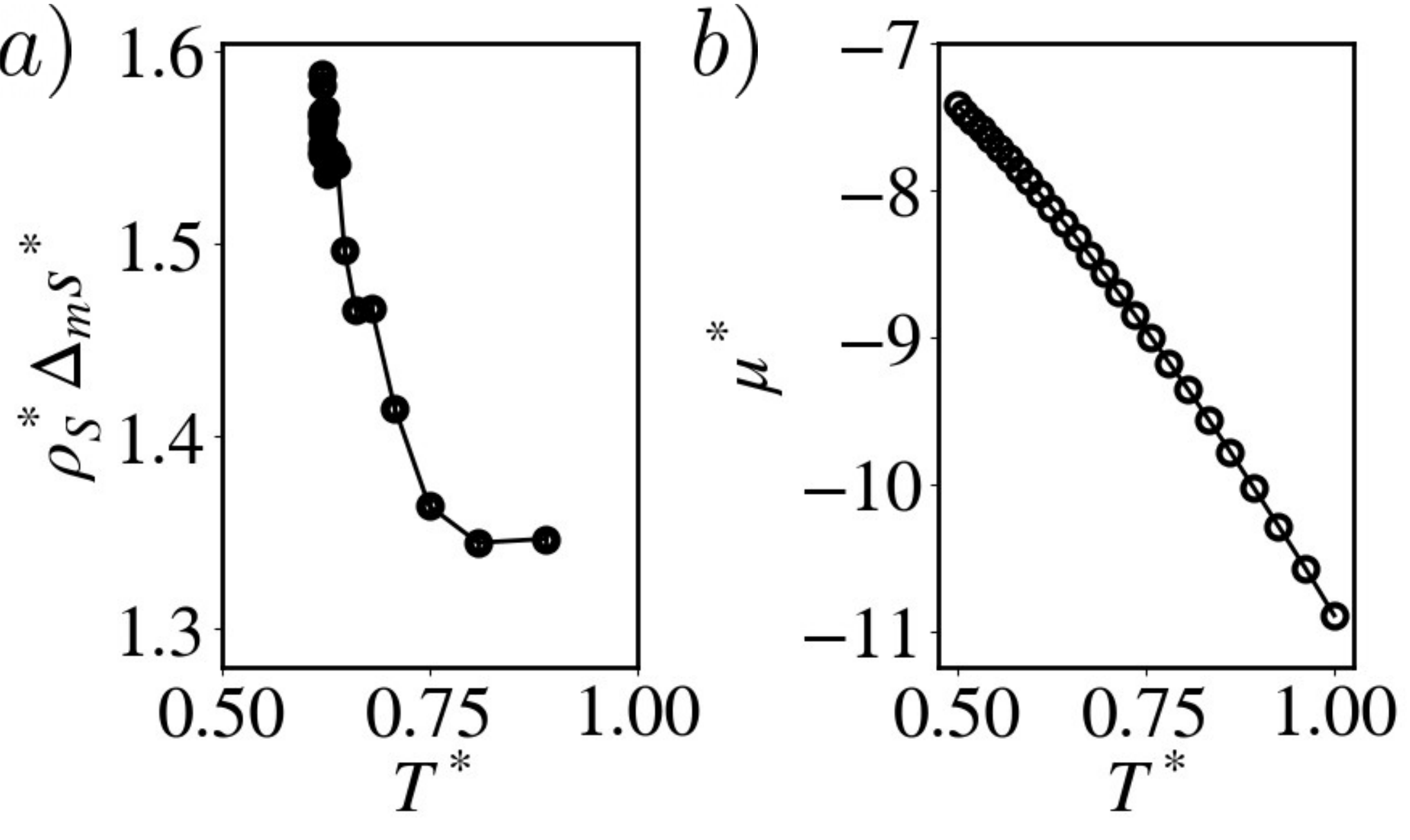}
    \caption{
    (a) Product of the bulk solid density $\rho_S^*$ with the bulk melting entropy per particle $\Delta_m s^*$ along the liquid-solid coexistence line and (b) chemical potential $\mu^*$ as a function of temperature, along the liquid-vapor coexistence line (see Fig.~\ref{fig:bulkphd}).
    All thermodynamic quantities are in LJ units.
    }
    \label{fig:bulkprop}
\end{figure}

The quantity which enters in the denominator in Eq.~\ref{eq::gtT} is in fact the product of the bulk solid density with the bulk melting entropy per particle. The latter can be determined from the GDI simulations, which provide the enthalpy of the coexisting solid and liquid phases, hence $\Delta_m s=\Delta_m h / T_m$. 
Fig.~\ref{fig:bulkprop}a shows the product $\rho_S\Delta_m s$, as a function of temperature, along the liquid-solid coexistence line. The GT prediction relies on the assumption that one can use the value for the bulk coexistence, $\rho_S^b\Delta_m s^b$, corresponding in the present case to the triple point with $T_T^* = 0.62$. This resulting error is of only $\approx 5\%$ for $T^*=0.65$ but already $\approx 13\%$ for $T^*=0.75$.
Finally, Fig.~\ref{fig:bulkprop}b reports the chemical potential determined by Widom insertion\cite{widom_topics_1963} as a function of temperature along the liquid-vapor coexistence line. These values are necessary for the HPT-GCMC simulations of Section~\ref{sec:hptgcmc}, but will not be further commented here.


\section{Crystallization under confinement: surface tension difference}\label{sec:TI}

The last term in the GT equation that needs to be computed in order to predict the temperature shift induced by confinement is the surface tension difference $\gamma_{LW} - \gamma_{SW}$, which is positive if the walls favor the solid phase with respect to the liquid phase and negative otherwise.
Computing surface tensions can be done following either a mechanical route through the stress tensor\cite{irving_kirkwood,kirkwood_buff} or a thermodynamical approach, which uses the definition of the surface tension as a (Gibbs) free energy per surface area
\begin{equation}
\gamma = \left( \frac{\partial F}{\partial A_W} \right)_{NVT} = \left( \frac{\partial G}{\partial A_W} \right)_{NPT}
\, .
\end{equation}

In order to avoid difficulties of the mechanical route for solid-solid interfaces\cite{nijmeijer_microscopic_1990}, we use a thermodynamic integration procedure to obtain $\gamma_{LW} - \gamma_{SW}$ from the Gibbs free energy difference between the walls in contact with the liquid or the solid phase. To that end, a bias acting on the system is introduced, providing a handle to drive the phase transition, and the relevant thermodynamic quantities are computed to obtain the properties of the unbiased system. Here we use the collective variable $Q_6$, derived from the sixth order Steinhardt parameters\cite{steinhardt_bond-orientational_1983,auer_numerical_2005,lechner_accurate_2008,reinhardt_local_2012,rein_ten_wolde_numerical_1996,kawasaki_construction_2011,sanz_homogeneous_2013} defined in Appendix~\ref{app:cvq6}, which quantifies the average (over the system) local order: a large (resp. low) value corresponds to an ordered solid (resp. disordered fluid).

Such a procedure is computationally more demanding than the study of the bulk properties and cannot be performed systematically as a function of thermodynamic conditions. Therefore, we first identify suitable conditions in which both the confined liquid and solid phases are sufficiently metastable, \emph{i.e.} close to the coexistence line under confinement, which is not known \emph{a priori} (see also Section~\ref{sec:hptgcmc}). For example, at the bulk coexistence temperature and pressure the confined liquid tends to recrystallize, which points to an increase in the melting temperature under confinement.
We fix the pressure to $P^*=0.2036$ (which corresponds to a bulk melting temperature $T_m^*=0.638$) and perform $NPT$ simulations during which the temperature is slowly increased then decreased. The evolution of the system density during these temperature ramps exhibits a pronounced hysteresis pointing to the metastability of both liquid and solid phases over a finite temperature range, which is then confirmed by long simulations (10~ns) of the two phases at the selected temperature $T_{TI}^* = 0.659$. These unbiased simulations also allow to determine the characteristic values for the collective variable in the solid and liquid phases ($Q_6^S = 0.3632$ and $Q_6^L = 0.1425$ from 2~ns simulations at $T_{TI}^*$), which are then used in the definition of the bias.

The details of the thermodynamic integration are given in Appendix~\ref{app:TI-bias}. In a nutshell, it follows a three-step scheme
\begin{gather*}
\text{Biased solid } 
\xrightarrow[\textcolor{blue}{\textstyle \lambda_{0\rightarrow1}}]{\text{\textcolor{blue}{\normalsize 2. shift bias}}} 
\text{ Biased liquid} \\
\textcolor{blue}{\text{1. introduce bias}} \Bigg\uparrow \textcolor{blue}{\textstyle \alpha_{0\rightarrow1}}
\qquad \qquad \quad
\textcolor{blue}{\textstyle \alpha_{1\rightarrow0}} \Bigg\downarrow \textcolor{blue}{\text{3. remove bias \quad}}\\
\text{Non-biased solid } \quad  
\xrightarrow[]{\textcolor{red}{\textstyle \Delta G^{TI}}} \quad 
\text{Non-biased liquid} 
\end{gather*}
in which a bias on the $Q_6$ collective variable is introduced (via a parameter $\alpha$ rising from 0 to 1), shifted from the solid to the liquid (via a parameter $\lambda$ from 0 to 1) and removed (by decreasing $\alpha$ from 1 to 0). Along the way, one computes the relevant thermodynamic quantities, which are then integrated over the whole thermodynamic path (see Eq.~\ref{eq:dgti}) to obtain the Gibbs free energy $\Delta G^{TI}$ associated with this transformation between the non-biased confined solid and liquid phases
\begin{multline}\label{eq:dgti}
\Delta G^{TI} = \int \limits_0^1 d\alpha \left< \frac{\partial U}{\partial \alpha}  \right>_{NPT;\alpha, \lambda=0} + \int \limits_0^1 d\lambda \left< \frac{\partial U}{\partial \lambda}  \right>_{NPT;\alpha=1, \lambda} \\
- \int \limits_0^1 d\alpha \left< \frac{\partial U}{\partial \alpha}  \right>_{NPT;\alpha, \lambda=1}
\end{multline}

\begin{figure}[hbt!]
    \centering
    \includegraphics[width=1.\linewidth]{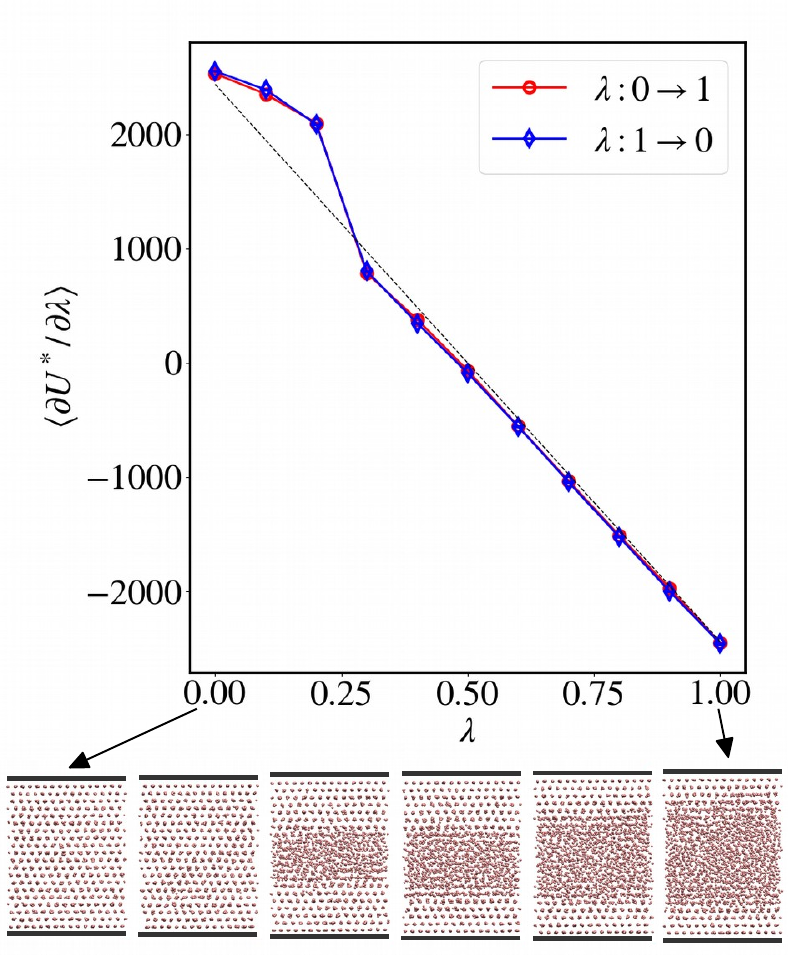}
    \caption{
    Thermodynamic integration to force phase transition. Energy derivative $\left\langle\partial U^* / \partial \lambda\right\rangle$ as a function of the biasing variable $\lambda$ for the shift of the bias step (in LJ units). Values are given for both the forward (solid-liquid, red open circles) and backward (liquid-solid, blue open diamonds) transformations. The black dotted line is a guide to the eye, with vanishing integral. Typical snapshots for several $\lambda$ values are shown, where LJ particles are in pink and the position of the Steele surfaces is indicated by the solid black lines.}
    \label{fig:ti-q6}
\end{figure}

Since the first and third step introduce/remove a bias of $Q_6$ towards the respective equilibrium values $Q_6^S$ and $Q_6^L$, their contributions to $\Delta G^{TI}$ are small (around 1 kJ/mol) and almost perfectly cancel each other. The final result then arises entirely from the intermediate step, which involves the derivative $\left\langle \partial U /\partial \lambda \right\rangle$, which is shown as a function of $\lambda$ in Fig.~\ref{fig:ti-q6}. In practice, the simulation for a given value of $\lambda$ is performed starting from a configuration obtained at a previous value. This may lead to hysteresis since the path (in configuration space) followed during the phase transition may differ in the forward (solid to liquid) and reverse (liquid to solid) processes. This is why alternative methods avoiding the explicit transition (such as the Frenkel-Ladd method\cite{frenkel_new_1984}, which involves known reference states such as the Einstein crystal and the ideal gas) are generally used\cite{das_effect_2013,mastny_melting_2007,grochola_constrained_2004,eike_toward_2004,wilding_freezing_2000,mcneil-watson_freezing_2006}. The results of Fig.~\ref{fig:ti-q6} show that with our choice of thermodynamic path, bias and simulation parameters (see Appendix~\ref{app:TI-bias} for more details), we achieve a good reversibility of the transformation.
The final result for the Gibbs free energy difference at $T_{TI}^*$ and $P^*$, taking into account the discretization error for the numerical integrations in Eq.~\ref{eq:dgti}, is $\Delta G^{TI*} (T_{TI}^*) = 50\pm28$ (in LJ units). This large uncertainty arises mainly from the numerical estimate of the integral, due to the jump between $\lambda=0.3$ and 0.4. 

The Gibbs free energy difference $\Delta G^{TI}$ obtained from the above thermodynamic integration can be decomposed into volume and surface contributions:
\begin{align}
\Delta G^{TI} &= 
(\Delta_m H - T \Delta_m S) + 2 A_W (\gamma_{LW} - \gamma_{SW})  \nonumber \\
&= \Gamma_{bulk} A_W \left(1 - \frac{T}{T_m^b}\right)\Delta_m h^b + 2 A_W (\gamma_{LW} - \gamma_{SW}) 
\label{eq:volsurf}
\end{align}
where we introduced $\Gamma_{bulk}=N/A_W-2\Gamma$ the number of ``bulk'' atoms per unit surface, with $\Gamma$ the excess number of atoms at each interface. The latter can be determined from the density profiles, as discussed in Appendix~\ref{app:TI-surftension}.
Eq.~\ref{eq:volsurf} then leads to $\Delta \gamma^* = \gamma_{LW}^* - \gamma_{SW}^* = 0.40 \pm 0.05$.

Before turning to the implications for the GT prediction, we note that for the present system the strong attraction between the particles and the wall ($\epsilon_{WF}^*\approx2.8$) results in a pronounced structuration in the vicinity of the surface, with several solid-like layers even between the wall and the liquid phase (see the snapshot for $\lambda=1$ in Fig.~\ref{fig:ti-q6} and the density profiles in Appendix~\ref{app:TI-surftension}). This has two important consequences. 
Firstly, this probably explains why the surface tension between the liquid and the wall (covered by a few solid-like layers) is larger than that between the solid and the wall, \emph{i.e.} the positive sign of $\Delta\gamma^*$. Secondly, the width of these solid-like films on both sides reduces the effective size of the bulk liquid and solid regions, assumed to be sufficiently large for the GT equation to apply.


\section{Crystallization under confinement: melting temperature}
\label{sec:hptgcmc}

In the previous sections, we computed the terms entering in the GT equation~\ref{eq::gtT} and investigated the temperature dependence of some of these terms. Here, we finally compare the resulting predictions of this equation to the melting temperature for our model system under confinement as a function of the pore size $H$ with results from Hyper-Parallel Tempering Grand Canonical Monte Carlo (HPT-GCMC) simulations. This technique, explained in detail in Appendix~\ref{app:ptgcmc}, runs parallel replicas at different temperatures, regularly spaced in $\beta=1/k_BT$, each replica being a GCMC simulation (in the $\mu A_WHT$ ensemble) in contact with a chemical reservoir. To model the setup described in Section~\ref{sec:GT} and Fig~\ref{fig:gibbs-thomson}, the imposed chemical potential $\mu$ is taken from the liquid-vapor coexistence at the replica's temperature (see Fig.~\ref{fig:bulkprop}b). The exchange between replicas improves the sampling of phase space. The confidence interval for the melting temperature $T_m$ is estimated for each pore size from the evolution of the average number of particles as a function of temperature, as explained in Appendix~\ref{app:ptgcmc}, also supported by visual inspection of the equilibrated configurations.

\begin{figure}[hbt!]
    \centering
    \includegraphics[width=1.\linewidth]{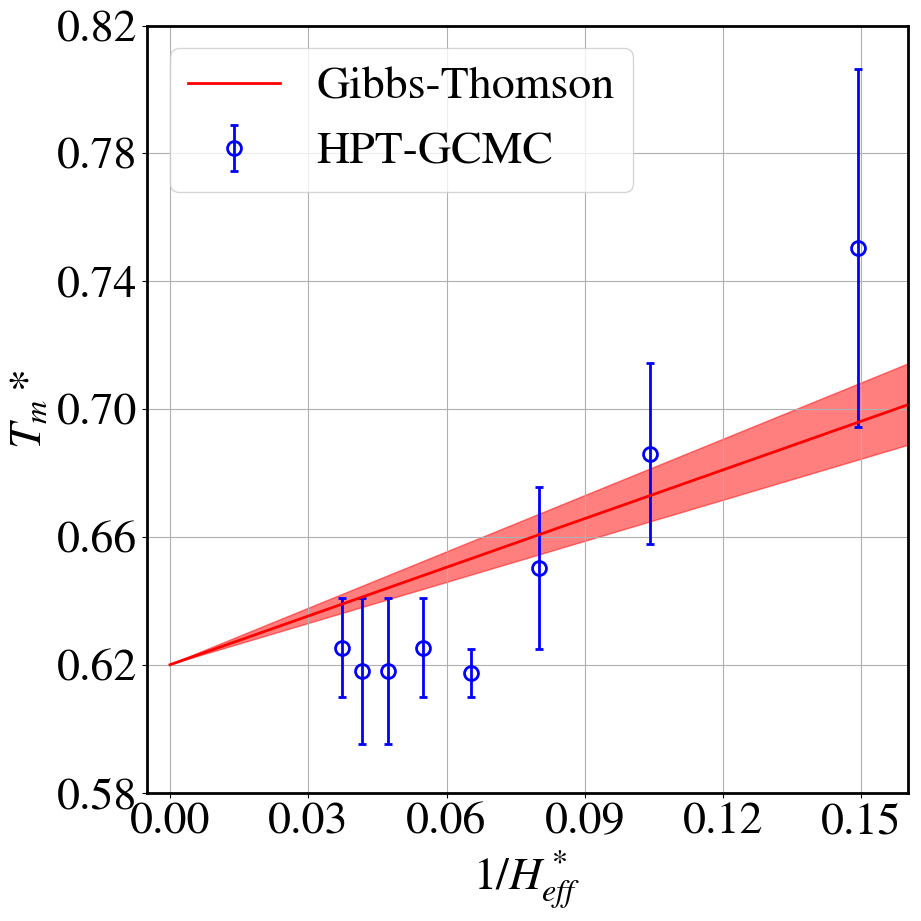}
    \caption{
    Melting temperature $T_m^*$ as a function of the inverse effective pore size $1/H_{eff}^*$ (see text). The red line indicates the prediction of the GT equation~\ref{eq::gtT} using the results of the previous sections (with the shaded area illustrating the uncertainty), while the open blue circles correspond to the direct determination of $T_m$ from HPT-GCMC simulations, together with their confidence interval (see text and Appendix~\ref{app:ptgcmc}).
    }
    \label{fig:TmvsInvH}
\end{figure}

In order to compare the results obtained by HPT-GCMC simulations to the prediction of the GT equation, one needs to consider the effective width of the pore occupied by the particles. From the position of the Gibbs dividing surfaces (see Appendix~\ref{app:TI-surftension}) located at $\approx\sigma$ from the Steele walls, we define $H_{eff}^* = H^*- 2$, with $H$ the distance between the positions of the walls (the difference between $H_{eff}^*$ and $H^*$ matters more in smaller pores, but does not influence the conclusions below). Fig.~\ref{fig:TmvsInvH} shows the melting temperature $T_m$ as a function of $1/H^*_{eff}$, together with the prediction of the GT equation using the results of Section~\ref{sec:bulk} for $T_T^b$ and $\rho_S\Delta_m s^b(T_T^b)$ and of Section~\ref{sec:TI} for $\Delta\gamma(T_{TI}^*)$.

One can first note that for large pores, despite the relatively large confidence interval due to the difficulties to converge the HPT-GCMC simulations which does not allow us to identify a trend with temperature, the results are consistent with the bulk value $T_T^b$ in the limit $H_{eff}\to\infty$. In addition, the order of magnitude of the GT predictions is consistent with the HPT-GCMC results down to very narrow pores (a few molecular diameters). However, the agreement is not quantitative, even for the larger pores considered in the present work ($\approx20$ molecular diameters). 

Importantly, though unsurprizingly, the GT equation fails to capture the transition from a regime dominated by the competition between volume and interfacial contributions, to a different one for small pores, dominated by disjoining pressure effects, \emph{i.e.} the mutual influence of the two interface. Even though this second regime is not the main focus of the present work and this is not visible with the considered pore sizes, the disjoining pressure oscillates due to the finite size of the particles and the formation of discrete layers at the interfaces, so that non-trivial effects on the thermodynamic behavior can be observed\cite{camara_molecular_2003,wan_confined_2012, das_effect_2013,kaneko_phase_2010,long_molecular_2013,kaneko_elevation/depression_2017}.

Several reasons can be put forward to explain the somewhat disappointing comparison between the GT prediction and the HPT-GCMC simulations for large pores. Firstly, there are uncertainties associated with the determination of the quantities entering the GT equation, but their combination does not seem too large in the large-pore regime. Secondly, the GT equation assumes that these quantities do not depend on the temperature or equivalently on the pore width. The results on $\rho_S^b\Delta_m s^b$ as a function of temperature in Section~\ref{sec:bulk} suggest that the effect would be limited to less than 10\% in the temperature range corresponding to large pores. Unfortunately, the other contribution to the GT slope, $\Delta\gamma$, could only be determined at a single temperature $T_{TI}^*$ (close to but different from $T_T^b$, to avoid the liquid-vapor coexistence), so that we cannot assess the effect of $T$ (or $H$) on the difference in surface tensions.

A further difficulty is that the chosen model system, with a dramatic ordering of the interfaces due to the strong attraction with the walls, leads to a small bulk region even for the larger pores considered here (see the density profiles in Appendix~\ref{app:TI-surftension}). This makes it particularly difficult to converge the HPT-GCMC simulations (the exchange of two replicas is unfavorable when the difference in the number of molecules, which increases with system size, is large) and generally increases the computational cost -- preventing \emph{e.g.} the systematic study of $\Delta\gamma$ with $T$ or $H$. One possibility to mitigate this difficulty would be to consider a different system with a weaker interaction with the walls, leading to only 2-3 layers at the interface -- more typical of simple liquids on flat walls than the 7-8 observed here. It is however not easy to predict the resulting effect on the magnitude of the temperature shift. Overall, the difficulties related to the sampling of crystallization under confinement suggest that evaluating the relevant quantities separately and using the GT equation may provide an interesting alternative route to predict the behavior in large pores from molecular simulations.


\section{Conclusion}

We revisited the derivation of the Gibbs-Thomson equation for the crystallization of a liquid confined in a slit pore, in order to clarify the definition of the system and corresponding thermodynamic ensemble, as well as the assumptions leading to the final result. We highlighted the importance of the thermodynamic conditions in the bulk reservoir in equilibrium with the confined system. We then tested the validity of the approximations by evaluating the physical quantities entering the GT equation (bulk density and melting entropy, difference in interfacial tensions) for a model system and, when possible, their evolution with the temperature. We finally compared the prediction of the GT equation, using these estimated properties, to the melting temperature obtained by HPT-GCMC of the confined system, as a function of the pore size. 

While the chosen model system turned out not to be ideal for this study, we found that the order of magnitude of the GT predictions is consistent with the simulations down to very narrow pores (a few molecular diameters), but is not quantitative even for the larger pores considered ($\approx20$ molecular diameters). Importantly, though unsurprizingly, the GT equation fails to capture the transition to a different regime for small pores, dominated by disjoining pressure effects, \emph{i.e.} the mutual influence of the two interfaces. Beyond the study of the GT equation, the present work highlights some difficulties related to the sampling of crystallization under confinement. Evaluating the relevant quantities separately and using the GT equation may provide an interesting alternative route to predict the behavior in large pores from molecular simulations, without resorting to computationally intensive techniques to determine the melting temperature for each confining length. 

The accuracy of the approximations leading to the GT equation depend of course on the nature of the fluid and of its interactions with the walls. However, the present approach to test them can be applied not only for model fluids such as the one considered here, but also more complex ones such as water or ionic liquids, provided that the relevant solid phases are known. When several solid phases need to be considered, the "confined Clapeyron" approach may not be efficient to explore the full phase diagram of the confined system. The GT equation in fact also implicitly assumes that a single phase transition is relevant in the range of considered thermodynamic conditions.

One could further use molecular simulations to go beyond some of the assumptions leading to the standard GT equation. For example, the temperature dependence of the density, melting entropy and difference in surface tensions could be explicitly included in the integral along the thermodynamic path connecting the bulk and confined systems. The evolution of $\Delta\gamma$ with temperature remains however computationally more demanding than that of the bulk properties reported here. Another direction for future work is to investigate other thermodynamic conditions in the reservoir. The case considered here corresponds to recent experiments on the capillary freezing of ionic liquids between the tip of an AFM and a substrate in mind\cite{comtet_nanoscale_2017}, but the extension to other conditions or ensemble is straightforward. For this particular system, we will also need to consider more realistic models of the liquid and of the substrate, including the effect of its metallicity\cite{scalfi_semiclassical_2020,scalfi_molecular_2021}. Of particular interest in this context is also the fact that the crystallization of confined fluids may also depends on the presence of an electric field\cite{zaragoza_phase_2018}. Finally, the coupling between phase transitions under confinement and mechanical properties~\cite{brochard_revisiting_2020} could similarly be investigated by combining continuum thermodynamics with molecular simulations to compute the relevant quantities.


\begin{acknowledgments} 
The authors are grateful to Lyd\'eric Bocquet for discussions on the nanoscale capillary freezing of ionic liquids and to Fabio Pietrucci and Guillaume Jeanmairet for their help with the PLUMED package. This project has received funding from the European Research Council  under the European Union's Horizon 2020 research and innovation programme (grant agreement No. 863473). The authors acknowledge HPC resources granted by GENCI (resources of CINES, Grant No A0070911054).
\end{acknowledgments}


\appendix

\section{Calculation of $\mathrm{d}\mu/\mathrm{d}T$ for an external liquid-gas equilibrium}\label{app:dmudt}

In order to express the temperature dependence of the chemical potential, $\mathrm{d}\mu/\mathrm{d}T$, imposed by the liquid-gas coexistence in the bulk reservoir, we study the corresponding bulk system in the $\mu V T$ ensemble and consider the liquid-gas transition (the gas phase will be noted with the subscript $G$). The thermodynamic potential is the grand potential $\Omega =  U - TS - \mu N = -PV$.
Along the coexistence line, the grand potential is equal in the two phases, \emph{i.e.} $\Omega_G = \Omega_L$, and so are the associated variations, \emph{i.e.} $d\Omega_G = d\Omega_L$. From the expression of the grand potential, the former equality immediately leads to $P_G = P_L$, while the latter results in:
\begin{gather}
    -S_G dT - P_G dV - N_G d\mu = -S_L dT - P_L dV - N_L d\mu \nonumber \quad.
\end{gather}
Using the equality of pressures, this yields
\begin{equation}
    \frac{\mathrm{d}\mu}{\mathrm{d}T} = - \frac{\rho_L s_L-\rho_G s_G}{\rho_L - \rho_G} \label{eq::dmudt} \quad.
\end{equation}
Eq.~\ref{eq::dmudt} relates the variations of the chemical potential $\mu$ to the variations of temperature $T$ at the liquid-gas coexistence.

\begin{figure}[hbt!]
    \centering
    \includegraphics[width=1.\linewidth]{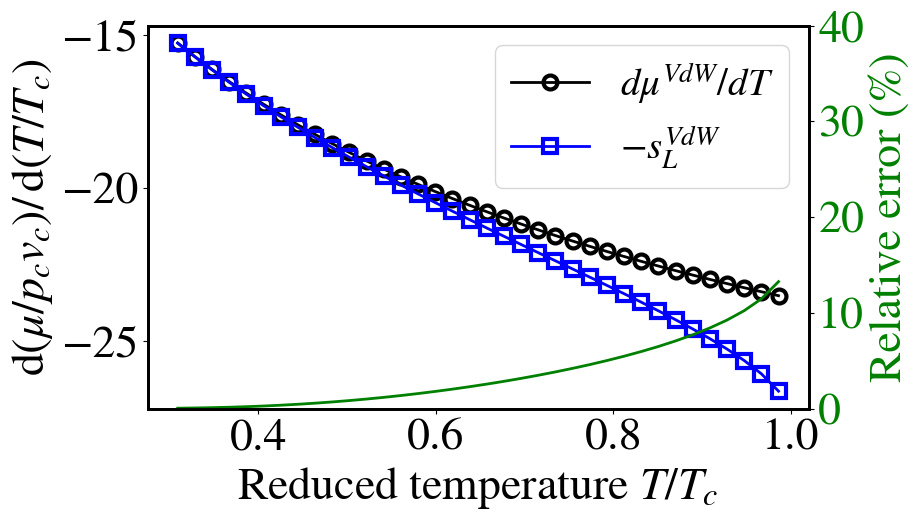}
    \caption{Evolution of $\mathrm{d}\mu/\mathrm{d}T$ as a function of temperature in reduced units for a van der Waals fluid ($\mu_{red}=\mu/p_cv_c$ et $T_{red}=T/T_c$, where the $c$ subscript refers to the critical point). Values of $\mathrm{d}\mu/\mathrm{d}T$ computed by Eq.~\ref{eq::dmudt} (open black circles) are compared to $-s_L$ using Eq.~\ref{eq::sL} (open blue squares). The green solid line is the relative error corresponding to approximating the former by the latter (Eq.~\ref{eq:approx}), with values indicated on the right y-axis. 
    }
    \label{fig:dmudT}
\end{figure}

In general, one can expect the density of the liquid to be larger than that of the gas ($\rho_L \gg \rho_G$) and the entropy per particle to be larger for the gas compared to the liquid ($s_G \gg s_L$). In order to make further progress, we consider the well-known van der Waals fluid, which is a good approximation to both the gas and the liquid phases, and use its equation of state, which amounts to a modified ideal gas law including an excluded volume $b$ and an attractive term $a$. A review of the ideal gas and van der Waals fluid properties is given in Ref.~\citenum{johnston_thermodynamic_2014} and gives the entropy per particle of the van der Waals fluid as
\begin{equation} \label{eq::sL}
    \frac{s}{k_B} = \ln \left( \frac{1 - b\rho}{ \rho\lambda^3} \right) + \frac{5}{2},
\end{equation}
where $\rho$ is the density and $\lambda ^3$ the quantum volume with $\lambda$ the De Broglie wavelength.
We introduce $\epsilon_\rho = \rho_G / \rho_L \ll 1$ and estimate the ratio $\rho_G s_G$ over $\rho_L s_L$ as
\begin{align}
    \frac{\rho_G s_G}{\rho_L s_L} = \frac{\epsilon_\rho \left[\ln (1 - b\rho_G) - \ln (\rho_L \lambda{^3}) + 5 /2\right] - \epsilon_\rho \ln \epsilon_\rho }{\ln (1 - b\rho_L) - \ln(\rho_L \lambda{^3}) + 5/2}   \label{eq::approx1}
\end{align}
In the limit where $\epsilon_\rho \to 0$ and $b \rho_G \to 0$ while $\rho_L$ is large but fixed, we have $\rho_G s_G \ll \rho_L s_L$, so that Eq.~\ref{eq::dmudt} can be reasonably approximated as
\begin{equation}
    \frac{\mathrm{d}\mu}{\mathrm{d}T} \approx -s_L \quad . \label{eq::approx}
\end{equation}
A numerical test of this approximation is shown in Fig.~\ref{fig:dmudT} for a van der Waals fluid, using the coexistence properties given in Ref.~\citenum{johnston_thermodynamic_2014}, for $b = \sqrt{2} \sigma^3$ and $a/b = 5 \pi \epsilon / 9$. Results are expressed in reduced units with respect to the critical temperature $T_c = 8a/(27b)$, pressure $p_c = a/(27b^2)$ and volume $v_c = 3b$. Reduced quantities are given in Fig.~\ref{fig:dmudT} along with the relative error made using the approximation in Eq.~\ref{eq::approx}. Values show an excellent agreement for temperatures small with respect to $T_c$ and a relative error smaller than 10\% on the relevant temperature range.

\section{Bulk simulation details}
\label{app:GDI}
The Gibbs-Duhem Integration method (GDI) method was implemented using the Python interface to LAMMPS\cite{plimpton_fast_1995} which allowed running two instances in parallel and coupling them during the run. The integration of the Clausius-Clapeyron equation (Eq.~\ref{eq:clapeyron}) was done using the predictor-corrector procedure as described by Kofke \emph{et al.} in Ref.~\citenum{kofke_direct_1993} using steps in reciprocal temperature $d\beta = 0.01$ for the liquid-vapor curve and steps in pressure $d\ln P = -0.4$ for the vertical part of the solid-liquid one to minimize integration errors. Each iteration of the predictor-corrector procedure was 20 ps long (with a timestep of 2 fs) and after convergence equilibrated data for $\Delta h$ and $\Delta v$ were collected for 200 ps. Gibbs-Ensemble Monte Carlo simulations and Widom insertion method were also run using an in-house code based on the Python interface to LAMMPS\cite{plimpton_fast_1995} to compute the interactions.

\section{Collective variable $Q_6$}
\label{app:cvq6}
The collective variable considered in this work is based on the sixth order Steinhardt parameters, which allow to measure the degree of order in the first coordination shell of a given atom\cite{steinhardt_bond-orientational_1983,auer_numerical_2005,lechner_accurate_2008}. We use a continuous version of the Steinhardt parameter which allows to compute derivatives and is given for each atom $i$ as the complex vector
\begin{equation}
q_{6m}(i) = \frac{\sum \limits_j \sigma(r_{ij}) Y_{6m}(\mathbf{r}_{ij})}{\sum \limits_j \sigma(r_{ij})} \, ,
\end{equation}
where the sum is on all other atoms $j$, $Y_{6m}$ is one of the sixth order spherical harmonics, with $m \in \llbracket-6, 6\rrbracket$ and $\sigma(r_{ij})$ is a switching function that depends on the distance $r_{ij}$ between atoms $i$ and $j$ and goes smoothly from 1 to 0 at a cutoff distance of $1.32\sigma$, selecting only first-shell neighbours of atom $i$.

We obtain a collective variable $Q_6$ that characterizes the whole system by taking the norm of the average vector $\overline{\mathbf{q}_{6}}$ over all atoms
\begin{equation}
Q_6 (\{\mathbf{r}_i\}) = \sqrt{\sum \limits_{m=-6}^6 
 |\overline{q_{6m}}|^2 } \, .
\end{equation}
We used the implementation available in the crystallization module of PLUMED~\cite{bonomi_promoting_2019,tribello_plumed_2014}.

\section{Confined phase transition using thermodynamic integration}
\label{app:TI}

\subsection{Thermodynamic integration in the $NPT$ ensemble}\label{app:TI-NPT}

We perform the thermodynamic integration with respect to the control parameter $\lambda$, which changes the total energy $U(\mathbf{r}^N; \lambda)$ of the system, in the $NPT$ ensemble. The corresponding thermodynamic potential is the Gibbs free energy $G(N,P,T;\lambda) = - \beta^{-1} \ln \Delta(N,P,T,\lambda)$, with $\Delta$ the partition function of the isothermal-isobaric ensemble and $\beta = 1/k_BT$ is the inverse thermal energy. The derivative of $G$ with respect to $\lambda$ is
\begin{align}
    \frac{\partial G}{\partial \lambda}& (N,P,T;\lambda) = - \frac{\beta^{-1}}{\Delta(\lambda)} \frac{\partial \Delta(\lambda)}{\partial \lambda}&\nonumber\\
    &= \frac{1}{\Delta(\lambda)}  \frac{\beta P}{\Lambda^{3N}N!}\int dV \int d\mathbf{r}^N  \frac{\partial U}{\partial \lambda} e^{-\beta (U(\mathbf{r}^N; \lambda) + PV)}&\nonumber\\
    &= \left< \frac{\partial U}{\partial \lambda}  \right>_{NPT;\lambda}&
\end{align}
where the brakets denote an ensemble average at fixed $N$, $P$, $T$ and $\lambda$. The Gibbs free energy difference of interest can therefore be obtained as
\begin{gather}
\Delta G = G(\lambda=1) - G(\lambda=0) = \int \limits_0^1 d\lambda \left< \frac{\partial U}{\partial \lambda}  \right>_{NPT;\lambda}
\; .
\end{gather}

\subsection{Thermodynamic path}\label{app:TI-bias}

The Hamiltonian $U_0$ is modified by introducing a biasing potential acting on the collective variable $Q_6$
\begin{equation}
U_{\text{bias}}(\lambda) = \frac{k}{2} \left[(1-\lambda) (Q_6 - Q_6^S)^2 + \lambda (Q_6 - Q_6^L)^2 \right]
\end{equation}
that will trigger the phase transition.
The initial state is the (free) solid phase and the final state is the (free) liquid phase. The thermodynamic integration procedure is divided into three steps:
\begin{enumerate}
\item Introduction of the biasing potential: $\alpha \in \left[ 0, 1 \right]$, $\lambda = 0$
\begin{gather}
	U(\alpha) = U_0 + \alpha U_{\text{bias}}(\lambda = 0)\\
	\frac{\partial U(\alpha)}{\partial \alpha} = \frac{k}{2} (Q_6 - Q_6^S)^2
\end{gather}

\item Shift of the bias from the solid to the liquid phase:\\ $\alpha=1$, $\lambda \in \left[ 0, 1 \right]$
\begin{gather}
	U(\lambda) = U_0 + U_{\text{bias}}(\lambda) \\
    \frac{\partial U(\lambda)}{\partial \lambda} = \frac{k}{2} \left[(Q_6 - Q_6^L)^2 - (Q_6 - Q_6^S)^2 \right]	
\end{gather}

\item Destruction of the biasing potential: $\alpha \in \left[ 1, 0 \right]$, $\lambda = 1$
\begin{gather}
    U(\alpha) = U_0 + \alpha U_{\text{bias}}(\lambda=1)\\ 
    \frac{\partial U(\alpha)}{\partial \alpha} = \frac{k}{2} (Q_6 - Q_6^L)^2
\end{gather}
\end{enumerate}

We ran 6 points in $\alpha$ from 0 to 1 (in steps of 0.2) and 11 points in $\lambda$ from 0 to 1 (in steps of 0.1), using a spring constant $k = 10^5$~kJ/mol. The bias is applied using the open-source PLUMED library~\cite{bonomi_promoting_2019,tribello_plumed_2014} coupled to the simulation code MetalWalls\cite{marin-lafleche_metalwalls_2020}. Each $\alpha$ or $\lambda$ point was first equilibrated for at least 20 ps, then run for at least 400~ps. For $\lambda = 0.3$, a small hysteresis was observed, which could be cured by simulated annealing, \emph{i.e.} heating the system at $T^* = 0.751$ for 100~ps and cooling it back to $T_{TI}^*$.

\subsection{Surface excess}\label{app:TI-surftension}

\begin{figure}[hbt!]
\centering
\includegraphics[width=\linewidth]{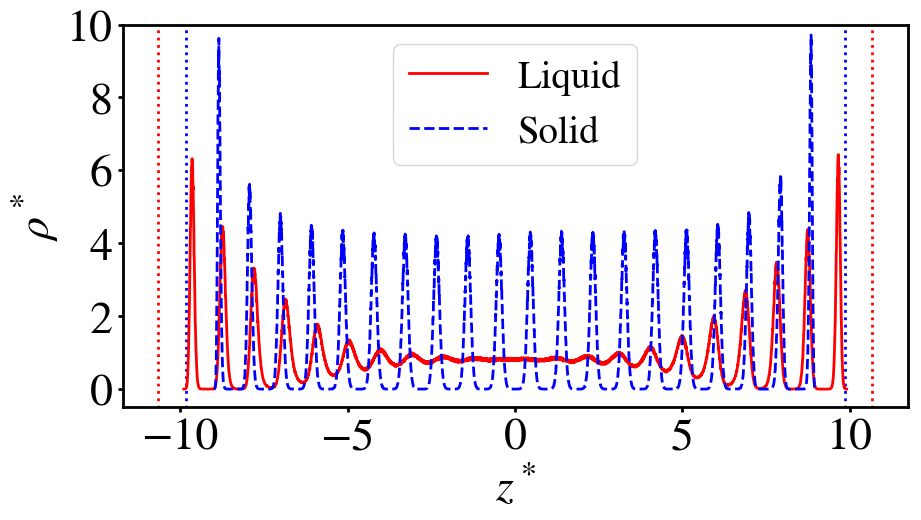}
\caption{
Density profiles across the pore, for the system described in Section~\ref{sec:TI}, for the liquid (red solid line) and the solid (blue dashed line) phases. The profiles were obtained from equilibrium $NPT$ simulations of each phase at $T_{TI}^*$; the average position of the walls is indicated by vertical dotted lines.
}
\label{fig:rho}
\end{figure}

The separation between volume and surface contributions to the Gibbs free energy $\Delta G^{TI}$ in Eq.~\ref{eq:volsurf} requires the computation of the surface excess $\Gamma$ at each interface (or equivalently the number $\Gamma_{bulk}$ of ``bulk'' atoms per unit surface of the system. This can be achieved from the density profiles $\rho(z)$ across the pore, shown for the liquid and solid phases in Fig.~\ref{fig:rho}. 

The thermodynamic definition of the surface excess is based on the position of the Gibbs dividing surface (GDS), $z_{GDS}$, which corresponds to an equivalent sharp interface between two homogeneous regions with densities $\rho_{wall}=0$ (in the wall) and $\rho_{bulk}=\rho_{L}$ or $\rho_{S}$ in the bulk region of the pore (averaged over a lattice spacing in the case of the solid phase):
\begin{equation}
\int_{z_{wall}}^{z_{GDS}} (\rho(z) - \rho_{wall}) dz
= \int_{z_{GDS}}^{z_{bulk}} (\rho(z) - \rho_{bulk}) dz
\end{equation}
with $z_{wall}$ and $z_{bulk}$ two positions in the wall and the bulk regions, respectively (we take $z_{bulk}=0$ in the center of the pore).
The bulk densities of the solid and liquid phases are $\rho_{S}^*= 0.936$ and $\rho_{L}^*= 0.826$. In practice, we find that the GDS is approximately located near the center of the first density peak, as expected. The surface excess can then be computed as $\Gamma=\int _{z_{GDS}}^{z_{bulk}} (\rho(z) - \rho_{bulk}) dz$, from which we obtain $\Gamma_{bulk}=N/A_W-2\Gamma$. Slightly different values of are obtained from the density profiles for the liquid and the solid phases. In the main text, we use the average and half difference for our final estimate of $\Gamma_{bulk}^*=17.0\pm0.2$ and its uncertainty.
Eq.~\ref{eq:volsurf} then leads straigthforwardly to the difference $\Delta\gamma=\gamma_{LW} - \gamma_{SW}$ from $\Delta G^{TI}$ and $\Gamma_{bulk}$. 

\section{Hyper-Parallel Tempering Grand Canonical Monte Carlo simulations (HPT-GCMC)}
\label{app:ptgcmc}

The hyper parallel tempering technique\cite{yan_hyperparallel_2000} is an extended version of the parallel tempering method in which replicas of the system at different thermodynamic conditions (e.g. temperature, pressure, chemical potential) are considered in parallel. This method can be extended to the Grand Canonical ensemble (constant volume, temperature, and chemical potential) to determine freezing and melting of a nanoconfined fluid in equilibrium with a bulk reservoir of the same fluid\cite{coasne_freezing_2007,coasne_effect_2009}. Each of the $M$ replicas consists of the Lennard-Jones fluid at a given set of temperature/chemical potentials [$T, \mu$] with $\mu(T)$ chosen to correspond to its value at the bulk liquid-gas phase coexistence (in practice, $M = 16$ was chosen in the present work). For each replica, conventional Monte Carlo moves in the Grand Canonical ensemble are carried out (particle translation, deletion and insertion). In addition, swap moves between a configuration 1 (energy $U_1$, $N_1$ particles) in replica A and configuration 2 (energy $U_2$, $N_2$ particles) in replica B are attempted. Swapping is accepted or rejected according to the following Metropolis probability
\begin{equation}\label{eq:pacc}
P_{acc} (\textrm{A}_1,\textrm{B}_2 \rightarrow \textrm{A}_2,\textrm{B}_1) = \min \left\{ 1, \frac{\rho_\textrm{A}(U_2,N_2)\rho_\textrm{B}(U_1,N_1)}{\rho_\textrm{A}(U_1,N_1) \rho_\textrm{B}(U_2,N_2)} \right\}
\end{equation}
where $\rho_\textrm{A}(U,N) \sim V^N/\Lambda_\textrm{A}^{3N} N! \times \exp[-\beta_\textrm{A}(U - \mu_\textrm{A} N)]$ and $\rho_\textrm{B}(U,N) \sim V^N/\Lambda_\textrm{B}^{3N} N! \times \exp[-\beta_\textrm{B}(U - \mu_\textrm{B} N)]$ are the density of states in the Grand Canonical ensemble for a system having a constant volume taken at $[T_\textrm{A}$, $\mu_\textrm{A}]$ and $[T_\textrm{B}$, $\mu_\textrm{B}]$, respectively. In these expressions, $\beta = 1/k_\textrm{B}T$ is the reciprocal thermal energy, while $\Lambda_\textrm{A}$ and $\Lambda_\textrm{B}$ are De Broglie thermal wavelengths at temperatures $T_\textrm{A}$ and $T_\textrm{B}$. Eq.~\ref{eq:pacc} then leads to
\begin{multline}
P_{acc}\left(\textrm{A}_1,\textrm{B}_2 \rightarrow \textrm{A}_2,\textrm{B}_1\right) 
= \min \{ 1, \left[  \frac{\Lambda_\textrm{A}}{\Lambda_\textrm{B}}  \right]^{3(N_1 - N_2)} \times\\
\exp[(\beta_\textrm{B} - \beta_\textrm{A})(U_2 - U_1) + (\beta_\textrm{B}\mu_\textrm{B} - \beta_\textrm{A}\mu_\textrm{A})(N_1 - N_2)] \}
\end{multline}

In this work, the temperature of the different replicas were chosen to correspond to a constant step in the reciprocal temperature $\beta$ between two successive replicas (roughly corresponding to a temperature step $\Delta T$ = 1--3 K). As shown in Ref.~\citenum{jin_molecular_2017}, HPT provides an accurate estimate of melting/freezing if both liquid and crystal configurations are considered in the initial replicas. In order to quantify the hysteresis between melting and freezing for each pore size, we performed two sets of simulations starting either from only crystal configurations or only liquid configurations. The convergence is monitored by following the number of atoms. Swapping between the crystal and liquid at different temperatures/chemical potentials improves the sampling of phase space, although it remains limited once the replicas have diverged in  terms  of  number  of  particles  and  energy. 

\begin{figure}[hbt!]
\centering
\includegraphics[width=\linewidth]{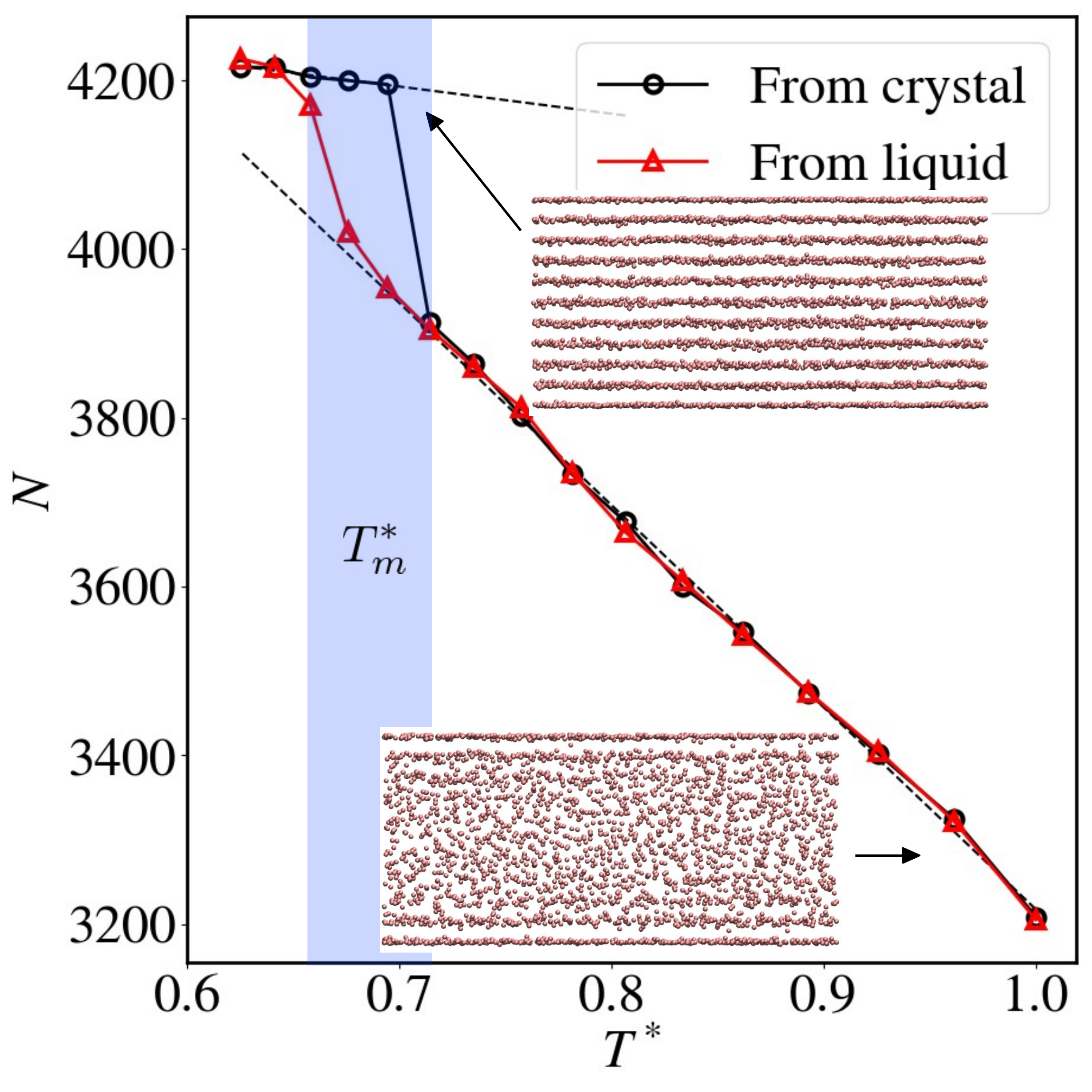}
\caption{
Average number of atoms $N$ in each HPT-GCMC replica as a function of the replica's temperature $T^*$ for a pore size $H^* = 11.6$. The two sets of data are obtained starting either from crystal (black circles) or liquid (red triangles) configurations in all replicas. Black dashed lines are linear fits to the low and high temperature regions, which are used to locate the melting temperature $T_m$ (blue shaded area indicating the confidence interval). The snapshots illustrate typical  crystal (top right) and liquid (bottom) configurations.
}
\label{fig:NofT}
\end{figure}

A confidence interval for the melting temperature $T_m$ can be obtained by identifying the transition region between solid (at low $T$) and liquid (at high $T$) phases. To that end, we analyze the average number of atoms per replica for each pore size, as illustrated in Fig.~\ref{fig:NofT} for $H^* = 11.6$. The approximately linear evolution of the number $N$ with temperature in the low and high temperature regions corresponds to the thermal expansion of the solid and liquid phases (also illustrated by typical snapshots in Fig.~\ref{fig:NofT}), respectively. Even though these two regimes are identical in the two sets of simulations starting from only crystal or only liquid configurations, we observe a hysteresis in the transition region. The corresponding range of temperatures is used as our confidence interval for $T_m$, reported for all pore sizes in Fig.~\ref{fig:TmvsInvH}.

\section*{DATA AVAILABILITY}

The data that support the findings of this study are available from the corresponding author upon reasonable request. This article may be downloaded for personal use only. Any other use requires prior permission of the author and AIP Publishing. This article appeared in \textit{The Journal of Chemical Physics} 2021, 154, 114711, and may be found at \url{https://aip.scitation.org/doi/10.1063/5.0044330}.


%

\end{document}